\documentclass{aa}
\usepackage{graphicx}
\usepackage{natbib}
\usepackage[cmex10]{amsmath}
\bibpunct{(}{)}{;}{a}{}{,} 
\begin{document}
%
\def\reffont{}
\def\reffonta{}
\title{Pixelization and Dynamic Range in Radio Interferometry}
\authorrunning{Cotton \& Uson}

  \author{ W.~D.~Cotton
  \and Juan M. Uson
	}

  \institute{National Radio Astronomy 
	Observatory\thanks{The National Radio Astronomy Observatory
		(NRAO) is operated by Associated Universities Inc.,
		under cooperative agreement with the National Science 
                Foundation.}, 
	520 Edgemont Road, Charlottesville, VA 22903-2475, USA.
         }
   \offprints{W.~D.~Cotton}

   \date{Received 19 November 2007 ; accepted 5 September 2008 }

   \abstract{
 This study investigates some of the consequences of representing the sky
by a rectangular grid of pixels on the dynamic range of images derived
from radio interferometric measurements.
In particular, the effects of image pixelization coupled to the CLEAN
deconvolution representation of the sky as a set of discrete delta
functions can limit the dynamic range obtained when representing
bright emission not confined to pixels on the grid.
{\reffont  Sky curvature effects on
non-coplanar arrays will limit the dynamic range even if strong sources are
centered on a pixel in a ``fly's eye'' representation when such pixel is not
located at the corresponding facet's tangent point.  Uncertainties in the
response function of the individual antennas as well as in the calibration
of actual data due to ionospheric, atmospheric or other effects will limit
the dynamic range even when using grid-less subtraction (i.e. in the
visibility domain)
of strong sources located within the field of view of the observation. }
A technique to reduce these effects is described and examples from an
implementation in the Obit package are given.
{\reffonta Application of this technique leads to significantly
superior results without a significant increase in the computing time.
}
   }

   \maketitle
%

\section{Introduction}
With the new generation of high sensitivity
interferometers to come on-line in the next few years (EVLA, ALMA,
LOFAR) wide--field imaging will be necessary to achieve the
sensitivity possible with these instruments. 
The problem is especially acute at low frequencies ($<10$ GHz)
where every field of view will contain several relatively bright
sources at any time.
The sensitivity of instruments such as LOFAR or the EVLA at lower
frequencies may be compromised much of the time by artifacts due to the
bright sources in the field if these artifacts are not reduced to an
acceptable level.
This paper describes artifacts arising from using pixelated images to
describe the sky as well as a technique for reducing them.
All data manipulations discussed in this report used the Obit package
(\cite{OBIT}, http://www.cv.nrao.edu/$\sim$bcotton/Obit.html).

\section {Effects of Pixelated Images}
It is generally convenient to represent the sky seen by an imaging
interferometer as a set of pixel values on a rectangular grid.
This is a good match to the widely used CLEAN deconvolution technique
which represents the sky as a set of delta functions located at the
centers of cells on such a grid.
A commonly used measure of the quality of an image is its ``dynamic
range,'' generally defined to be the ratio of the brightest pixel in
an image to the RMS pixel-to-pixel fluctuation in areas devoid of
emission. 
{\reffonta 
Application of this criterion is generally straightforward as the
response of the primary beam leads to mostly empty regions in the
images sufficiently far from the pointing center.
This convention is adopted in the following.}

\begin{figure*}[!t]
\centerline{
\includegraphics[height=3.5in,angle=-90]{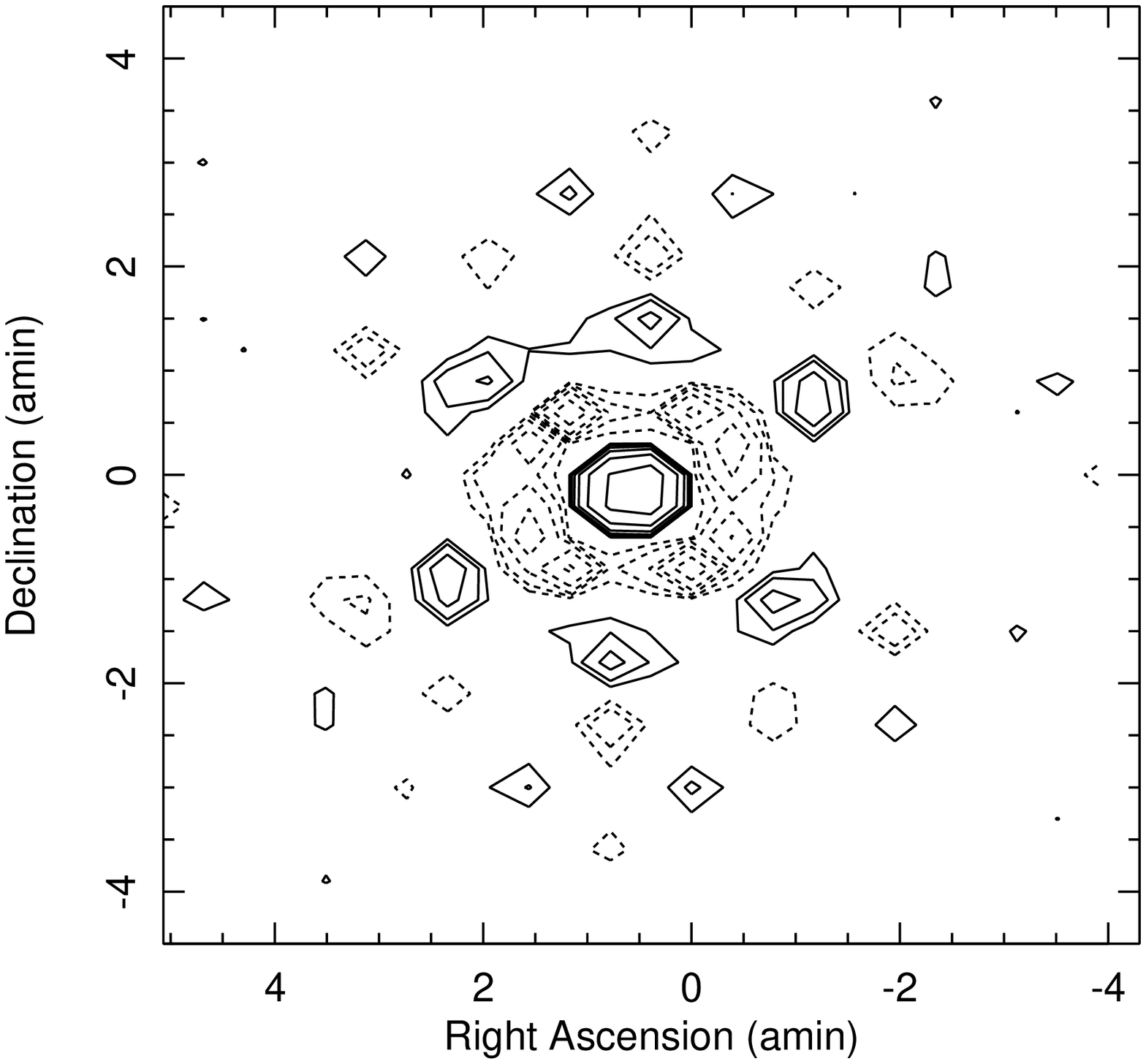}
\includegraphics[height=3.5in,angle=-90]{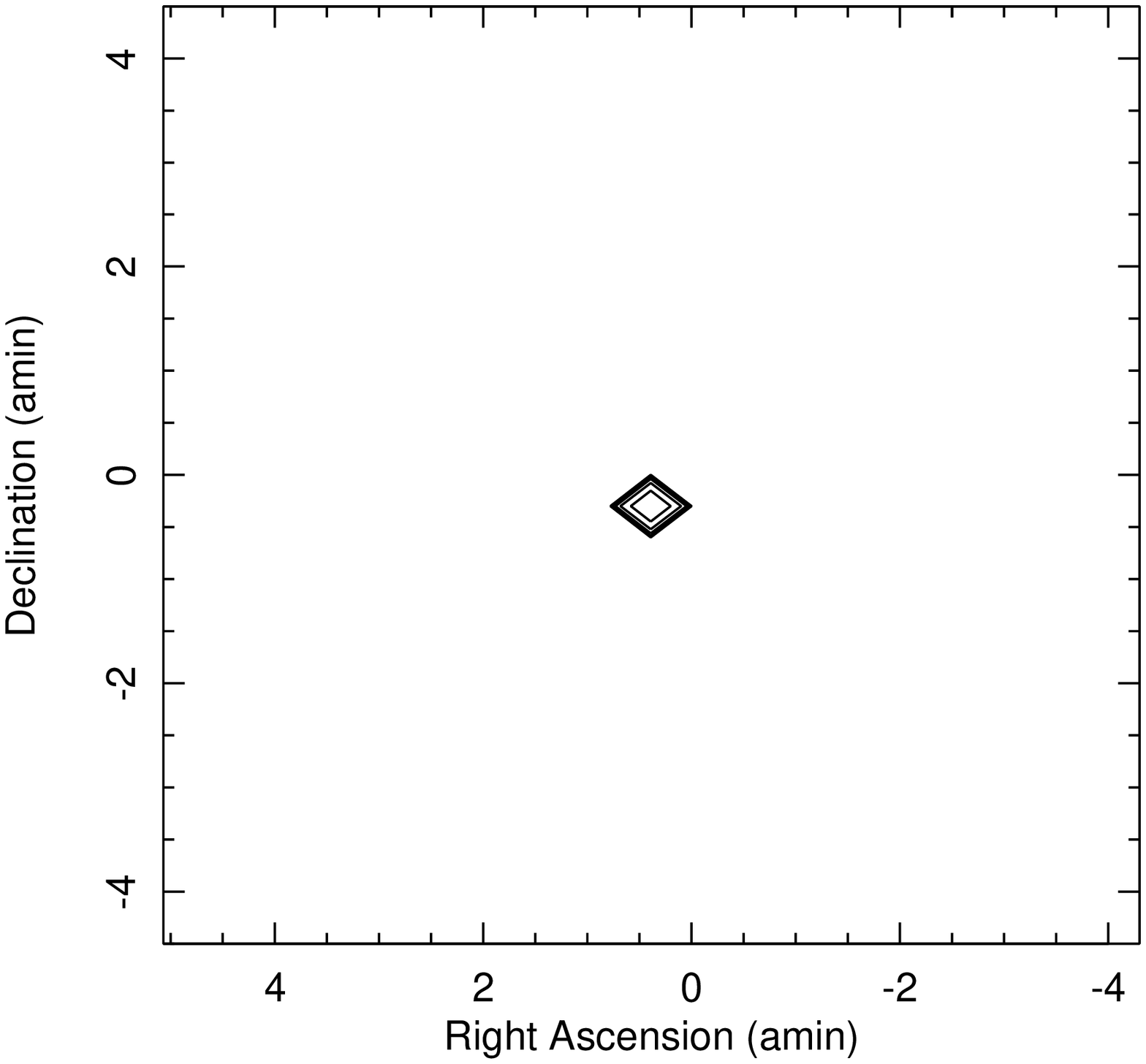}
}
\caption{ {\reffonta
Contour plots of example CLEAN images derived from noiseless point
source model data restored using delta functions for the components to
demonstrate the actual distribution of locations of CLEAN components.
The plots show the same region and have the same contour levels,
factors of powers of 2 x 0.1\% of the model flux density; negative
contours are dashed.
\hfil\break
{\bf Left:} The point source was located approximately midway between cells in
both dimensions.
CLEANing used 1000 components.
\hfil\break
{\bf Right:} The point source was located exactly on the central
pixel.
Only the central pixel contains emission; CLEANing used 200 components.
}
}
\label{Example}
\end{figure*}

One limitation of the pixelization technique is that emission not
confined to points on the grid cannot be represented exactly and the
CLEAN technique will approximate such structure by a potentially
infinite series of alternating positive and negative delta functions.
The problem is particularly severe when the image contains very bright,
unresolved emission as is common in the radio sky. 
A combination of the limited support of the actual representation and
the effects of the finite precision of digital computers will limit
the dynamic range obtained by introducing artifacts in the derived
image. 
{\reffonta
This effect has long been recognized as a problem \citep{Briggs1992,
Perley1999B}.
\citet{Briggs1992} describe the result as due to the discontinuity in
the visibility function of an off--center point source at the boundary
of the sampled region of the u--v plane which requires an infinite
number of components in the image plane to model it accurately.
\citet{Briggs1992} estimate that this effect will limit dynamic range
to $\approx 1000$.

This effect is easily understood for an {\reffont unresolved} source located
between pixels.
In order to model a point source between pixels, the deconvolution
must add emission in the adjacent pixels.
To counteract the resultant broadening of the source, negative
emission must then be added around the source.
An example of this effect is clearly shown in Figure
\ref{Example} which compares the results of CLEANing a model point
source both exactly centered on a pixel and offset between pixels.
For the offset source, the off--source RMS is 2.3 $\times 10^{-4}$
of the model source flux density (dynamic range = 4300) while for the
centered source, the off--source RMS is 4.7 $\times 10^{-10}$ of the
model source flux density (dynamic range = 2.1$\times 10^{9}$). 
In the latter case, the result is limited by the precision of 32--bit
digital arithmetic.
}

In the past, uncentered point sources have not been a particularly
serious problem as the bright source was usually the object under
investigation and centering it on a pixel was straightforward.
This is not the case for surveys or future observations where the
source(s) of interest may be faint sources in the presence of multiple,
much stronger sources whose locations on the imaging grid are not 
easily controlled.
{\reffonta
Processing of the NRAO VLA Sky Survey (NVSS) \citep{NVSS} used partial
pixel shifts to align the imaging grid to a single exceptionally
bright ($>0.5$ Jy) source whenever such a source was present in the field.
}

A second limitation of the pixelization technique is that the
rectangular grid is flat whereas the sky is not, see \cite{Cornwell1992}.
In the case of an array confined to a plane during synthesis, such as
the E-W linear Westerbork Synthesis Radio Telescope, a projection of
the sky is possible which avoids this problem; 
but, in the general case, this is not possible.
This effect is referred to in the following as the co-planarity problem.
If the field of interest is small enough, the curvature of the sky can
be negligible.
Several techniques have been developed to solve the more general
problem. 
Some of these are:
\begin{itemize}
\item Full 3-D imaging\\
Interferometric measurements are made in visibility space described by
the coordinate set (u,v,w).
These measurements can be convolved onto a 3D grid and Fourier
transformed into a 3D image.
The celestial sphere is a spherical surface in this 3D image.
A 3D deconvolution followed by projection onto a plane is possible but
in practice this is sufficiently expensive in computing resources that
it is not used.
See \cite{Perley1999A} for details.
\item Fly's Eye\\
The curved surface of the celestial sphere can be approximated by a
mosaic of facets, each tangent to the celestial sphere and of
sufficiently limited extent that the error introduced is
{\reffont negligible}.
However, the errors increase quadratically with the distance of a cell
to the corresponding tangent point and this can still limit the
dynamic range. 
See ``the Polyhedron Method'' in \cite{Cornwell1992} for details.
\item W projection\\
It is possible to correct for the diffractive effects on the
wavefront as it propagates from the antenna closer to the source to
the farther on each baseline.
This correction is made to the convolution of the visibility data onto
the grid prior to Fourier transformation.
This allows a single, flat 2D grid to represent the curved sky.
However, the derived ``dirty'' image is no longer strictly a convolution.
See \cite{Cornwell2005} for details.
\end{itemize}
In the following, only adaptations of the Fly's Eye technique are considered.

\section {Wide--field Imaging with Fly's Eye and Catalog of Sources}
The technique used for wide--field imaging in the following
tests is as follows.
A circular field of view to be completely imaged is specified by its
radius from the pointing center.
The data are examined and the cell spacing (if not specified) is
picked on the basis of the longest baseline in the data 
{(\reffonta one quarter of the smallest fringe spacing)}
and the size of a facet ``undistorted''  by co-planarity effects
determined from the maximum extent of any baseline in the selected
data in the direction of the pointing center.
{\reffonta
The radius of an undistorted region is adapted from \citet{Thompson99}
as $0.33\ \sqrt{1/maxW}$ radians where $maxW$ is the maximum value
of the ``w'' in the data set in wavelengths.
}
A ``Fly's Eye'' tessellation of 
{\reffonta circular }
regions in a hexagonal pattern is then defined which fully covers the
field of view and a set of square images enclosing these circular
regions defines a mosaic of facets. 

In general, the imaged field of view does not enclose all of the sky
to which the array elements have significant gain, so that 
sources are visible outside of this fully--imaged field of
view.
To include such sources, facets are added to the mosaic centered on
the positions of outlying sources obtained from a catalog (currently a
stripped--down version of the NVSS) which are deemed to
have an apparent brightness above a given 
{\reffonta user-specified}
threshold based on an
assumed spectral index and a model of the antenna gain pattern. 
{\reffonta  These additional facets do not need to be contiguous with
those covering the field of view wanted.
}
At high frequencies, a catalog could be generated from the WMAP
catalog of point sources \citep{WMAP}.
{\reffonta Accurate positions and flux densities are not required,
only a list of positions around which a bright source {\bf might}
appear, as the decision to auto--center is based on the results of the
initial CLEAN.}

The variant of CLEAN used in the following discussion is the ``visibility
based'' or ``Cotton-Schwab'' CLEAN \citep{Schwab1983, Cotton-SW89},
in which the Fourier transform of the estimate of
the sky is iteratively subtracted from the visibility data and the
residual image re-derived. 
This allows multiple independent ``facets'' to be imaged on the sky.
All facets in the mosaic to be deconvolved are formed and a quality
measure based on both peak (residual) brightness and extended
emission
({\reffonta 0.95 $\times$ peak absolute value residual in the clean window
  + 0.05 $\times$ the average residual - as used in the AIPS 
package\footnote{http://www.aips.nrao.edu}).
}
is used to determine which facet is to be CLEANed next.
Components are selected from this facet and then subtracted from the
visibility data.
The next highest quality measure facet is re-imaged and if it still
maintains its status, it is CLEANed, otherwise, the next highest
facet is re-imaged and tested, etc.
{\reffonta
The CLEAN is stopped when one of two conditions is satisfied, 1) the
total number of CLEAN components reaches a user specified limit or 2)
the maximum absolute value residual in the CLEAN window of all facets
is less than a user specified minimum, usually of order of the
anticipated RMS in the image.
}

After CLEAN is finished, components are (optionally) convolved
with a Gaussian approximation to the instrumental PSF and restored to
the facet from which they were subtracted as well as to any
overlapping facet containing their positions.
After restoration of the subtracted components, all facets are
projected (``flattened'') onto a grid covering the specified field of
view.

{\reffont
Because strong ``flanking sources'' can be observed with dedicated pointings
prior to reducing the data at hand and can thus be available in a catalog, it
would seem possible (in principle) to subtract them from the measured visibilities
prior to processing of the target field.  However, such visibility-based subtraction
is hindered by foreground effects on the calibration as well as by imprecise
knowledge of the response of the primary beam of the antennas which modify the
apparent position and flux density of the cataloged sources.  Efforts to fit these
effects in the uv-plane have had limited success.  The technique has been shown
to work reasonably well for only a few sources using simulated data and appears
to require vast computing resources to handle even a few sources \citep{Voronkov04}.  
}

\section {AutoCenter Technique}
In order to minimize pixelization related artifacts using the Fly's
Eye technique, it is desirable to locate each strong, point--like source
{\reffonta exactly}
on the pixel that is at the tangent point of the facet
that contains it.
{\reffonta This is accomplished by first identifying these sources in
an initial imaging step and then adding facets for strong sources not
already very close to a facet center.
}
It is also necessary to restrict CLEAN from assigning any components
to any overlapping regions of other facets.

An initial CLEAN is used to determine which objects in the field, if
any,  have sufficiently bright emission to warrant being centered on a 
special facet.
The CLEAN needs to be deep enough that an accurate measure of the
centroid of the source can be made to subsequently center it on a pixel.
{\reffont Currently,} the implementation in the Obit package CLEANs to
a factor of 0.1 of 
{\reffonta the user--specified auto--center brightness threshold
above which artifacts are expected to be above the noise level.}
The initial CLEAN level needed depends on the uv coverage, and dynamic
range but generally a factor of 100 to 1000 below the peak is adequate.
Also, since the residuals are not needed, it is not necessary to
derive a full set of residual images at the end of the initial CLEAN.
The decision 
{\reffonta that AutoCentering is needed }
is based on the highest pixel value in
the CLEAN-able portions of the initial dirty images.

Peak brightnesses are determined from the sum of the CLEAN components
within a given radius (cells within 62.5\% of the FWHM of the PSF in the
current implementation) of each component derived from each facet.
If this sum exceeds the {\reffonta auto--center brightness} threshold
level, then the centroid of the peak is determined from the CLEAN
components by a moment analysis.
A {\reffont small} facet, currently 96$\times$96 pixels centered at
the derived centroid, is added to the working imaging mosaic.
It is possible that the same source may appear in several overlapping
facets in the initial {\reffont fly's eye tiling} so it is necessary to ensure that a given
strong source {\reffont is added only once}, a coincidence of better than one-half of the
FWHM of the synthesized beam (typically two cells) is considered to
correspond to the same source.
If a source is already within 0.5 pixel of the tangent pixel of the
enclosing facet, a new facet is not created, but the facet is
re-centered (if not already within 0.01 cell).  

In order to ensure that CLEAN assigns no components to the re-centered
source appearing in overlapping facets of the initial mosaic, each of
the facets is examined to see if it contains the position of the
source to be re-centered. 
In any facet in which the re-centered source appears in the CLEAN-able
region, a round ``unbox,'' currently of radius 33 pixels, is added to
that facet centered on the position of the corresponding source.
This size is slightly smaller than the size of the initial cleanable
region in the new facet added, a radius of 38 pixels, to ensure
that no pixel will be excluded a priori from a component search {\reffont given the
non-coplanarity of the original and the new facets}.
These choices are somewhat arbitrary but seem to work well.
An ``unbox'' is like a normal CLEAN window except that any enclosed pixel
will not be considered as a location for CLEAN components
even if located inside of another regular CLEAN box,  i.e.
the CLEAN process ignores any pixels inside of an unbox.
Pixels inside of unboxes are also excluded from statistical estimates
such as maximum, minimum and RMS.
If any re--centering operations are required, the initial CLEAN is
repeated. 
After this re--centering, all prior CLEAN components are
discarded {\reffont before beginning a CLEAN}.

Because the accuracy of the determination of the centroid of a source
to be re-centered is adversely affected by the limited CLEAN and the
very effects this technique is trying to correct, some iteration may
be in order. 
Images in which high dynamic range is desired generally benefit
as well from one or more iterations of self--calibration.
At the beginning of each {\reffonta CLEAN},
the centroids of {\reffonta auto--centered sources from the previous CLEAN are}
checked to see if they are sufficiently close to the
tangent pixel.
If the first (reduced) moment of CLEAN components within 1.5 pixels of
the central pixel is offset by more than 0.01 pixel, then the
image is re-centered and all components from the previous CLEAN are
discarded. 
This allows an iterative refinement of the centroid of the peak and
improves significantly the results over a single estimate.

\section {Examples}
The following sections give two sets of examples processed using this
technique.
The first involves simulated data and the second, actual VLA
observations of 3C84.

\begin{table}[!t]
\centering
\caption{Offset Source Dynamic Ranges}
\vskip 0.1in
\begin{tabular}{|c|c|c|}   \hline
\hline
 Offset(pixels) &  DR$^a$  &  DR\_corr$^b$  \\
\hline
 0     &  $>1.0\times 10^{23}$   & $>1.0\times 10^{23}$  \\
 0.01  &  $260\times 10^{3}$     &   $9.2\times 10^{6}$  \\
 0.02  &  $130\times 10^{3}$     &   $10.1\times 10^{6}$ \\
 0.03  &  $94\times 10^{3}$    &   $10.3\times 10^{6}$ \\
 0.05  &  $60\times 10^{3}$    &   $8.9\times 10^{6}$  \\
 0.1   &  $29\times 10^{3}$    &   $8.9\times 10^{6}$  \\
 0.141 (0.1,0.1)$^c$ &  $17\times 10^{3}$    &   $4.9\times 10^{6}$  \\
 0.2   &  $17\times 10^{3}$    &   $11.5\times 10^{6}$ \\
 0.283 (0.2,0.2)$^c$ &  $9.7\times 10^{3}$     &   $4.4\times 10^{6}$  \\
 0.3   &  $12\times 10^{3}$    &   $10.2\times 10^{6}$ \\
 0.5   &  $10\times 10^{3}$    &   $9.6\times 10^{6}$  \\
 0.707 (0.5,0.5)$^c$ &  $6.2\times 10^{3}$    &   $7.1\times 10^{6}$  \\
\hline
 10    &  $65\times 10^{3}$    & $69.6\times 10^{6}$   \\
 20    &  $16\times 10^{3}$    & $79.1\times 10^{6}$   \\
 30    &  $7.3\times 10^{3}$    & $66.3\times 10^{6}$   \\
 50    &  $2.7\times 10^{3}$    & $8.1\times 10^{6}$    \\
 100   &  $0.74\times 10^{3}$    & $0.4\times 10^{6}$    \\
\hline
\end{tabular}
\hfill\break
Notes:\hfill\break
$^a$ Dynamic range without autoCenter\hfill\break
$^b$  Dynamic range with autoCenter\hfill\break
$^c$  Offsets along both coordinates\hfill\break
\label{pixel_offset}
\end{table}

\begin{figure}[!t]
\centering
\includegraphics[height=3.0in,angle=-90]{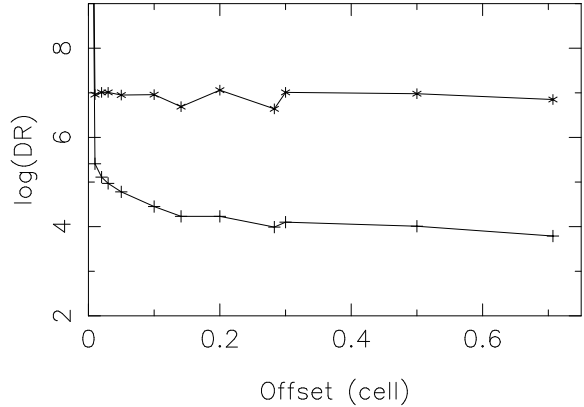}
\caption{ 
Dynamic Range (DR) obtained on simulated, noiseless data as a function of
fractional pixel offset. 
A combination of offsets on one and both axes are included.
The ``+'' symbols represent CLEANing without using autoCenter and
the ``$\ast$'' represent the results of using autoCenter.
}
\label{frac_pixel}
\end{figure}
\begin{figure}[!t]
\centering
\includegraphics[height=3.0in,angle=-90]{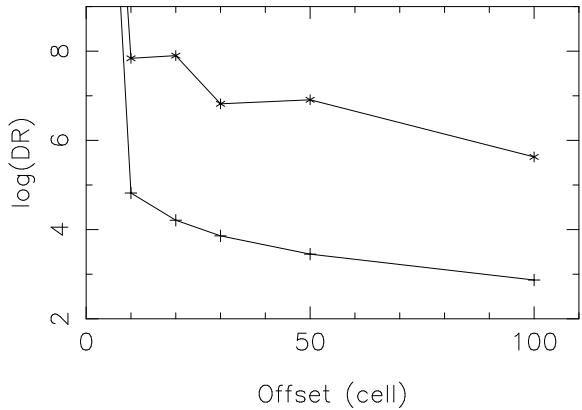}
\caption{ 
Dynamic Range (DR) obtained on simulated, noiseless data as a function of large,
whole pixel offset offsets to test non co-planarity effects.
The ``+'' symbols represent CLEANing without using autoCenter and
the ``$\ast$'' represent the results of using autoCenter.
}
\label{whole_pixel}
\end{figure}

\subsection {Simulated Data}
Simulated data have the advantage that their properties are known
which makes it simple to separate artifacts from source structure.
In these tests, the model used was a single 1 Jy point source with no noise
or other corruptions added.
In order to test the effects of fractional pixel offsets, the source
was located at the phase center of the data {\reffont and the data imaged with} a series
of fractional pixel shifts added to the center of the image.
The artificial data set used the same uv--plane sampling as a 74 MHz
VLA data set which consisted of 12 frequency channels of 122 kHz
bandwidth.  The tests were performed in the Obit package.  The CLEAN
used the visibility--based technique and proceeded for 1000 iterations
with a loop gain of 0.1 and a CLEAN window of radius 10 cells centered
on the peak.  
The model visibility computation used the  ``DFT'' method which is
more accurate than the ``GRID'' method as implemented in Obit (and
AIPS). 
Several iterations of imaging/re-centering were done in order to
refine the estimate of the centroid.
The cell spacing used was 0.25 of the FWHM of the fitted Gaussian
restoring beam so the image was reasonably over--sampled.
The components removed by the CLEAN procedure were not restored and
the RMS of the pixel values in the final residual image was used to
derive the dynamic range.
Each of the tests was then repeated turning on the autoCenter mode 
and the corresponding dynamic range {\reffont determined in turn}.
A combination of offsets on a single axis and on both axes are included.
The results are shown in Table \ref{pixel_offset} and Figure
\ref{frac_pixel}. 

A second set of tests explored the effects of non co--planarity by
inserting a series of large, whole--pixel offsets to the
position of the point source model and using a procedure like the one
described above for small pixel offsets.
The results are shown in Table \ref{pixel_offset} and Figure
\ref{whole_pixel}. 
All offsets except the one of 100 pixels are within the ``undistorted
field of view'' of a facet.

   While these tests are not exhaustive, it is clear that fractional
pixel offsets of bright point--like sources can limit the dynamic
range to  $\sim10^4$ and non co--planar effects can limit the
dynamic range to  $\sim10^3$ even with a moderately conservative
limit on facet size.
Applying the autoCenter technique improved the dynamic range by
typically a factor of 50 to 1000 for the fractional pixel offset tests
and typically by a factor of 1000 in the large pixel offset tests.

Figure \ref{frac_pixel} shows a fair amount of scatter in the
corrected dynamic range achieved.
We believe this to be the result of residual errors in the
centroiding as offsets on two axes produced lower dynamic range than
comparable offsets on a single axis.
All corrected dynamic range {\reffont values} were substantially better than that
obtained using a 0.01 pixel offset without correction.
At very high dynamic range the accuracy of the centroiding needs to
be exceedingly precise.
This is possible in this test as there are no systematic errors and
the model is exactly a point source.
In the real sky, resolution may be a problem even for a source whose
size is a very small fraction of the PSF; position errors of less
than 0.0025 of the PSF seem to be capable of limiting dynamic range so
resolution on similar scales might be a problem.
In Figure \ref{whole_pixel} the efficacy of the correction seems to
diminish with increasing pixel offset.
This may be as much a problem with the simulated data as with the imaging;
the larger position shifts needed to model a source with a substantial
offset from the pointing center will result in loss of numerical
precision.

   Non co--planar effects coupling to the pixelization can limit
significantly the dynamic range achieved.
A source observed 20\% of the way to the edge of its imaging facet
suffers comparable dynamic range loss to a central source observed 0.2
cells from the closest pixel.
This indicates that imaging using the Fly's Eye technique
needs to be applied with bright sources at the center of a facet
(tangent point) and not merely on a pixel.
A coarser grid spacing will likely lead to larger errors than
presented here.

\subsection {Actual Data}
\begin{figure*}[!t]
\centering
\includegraphics[height=3.5in]{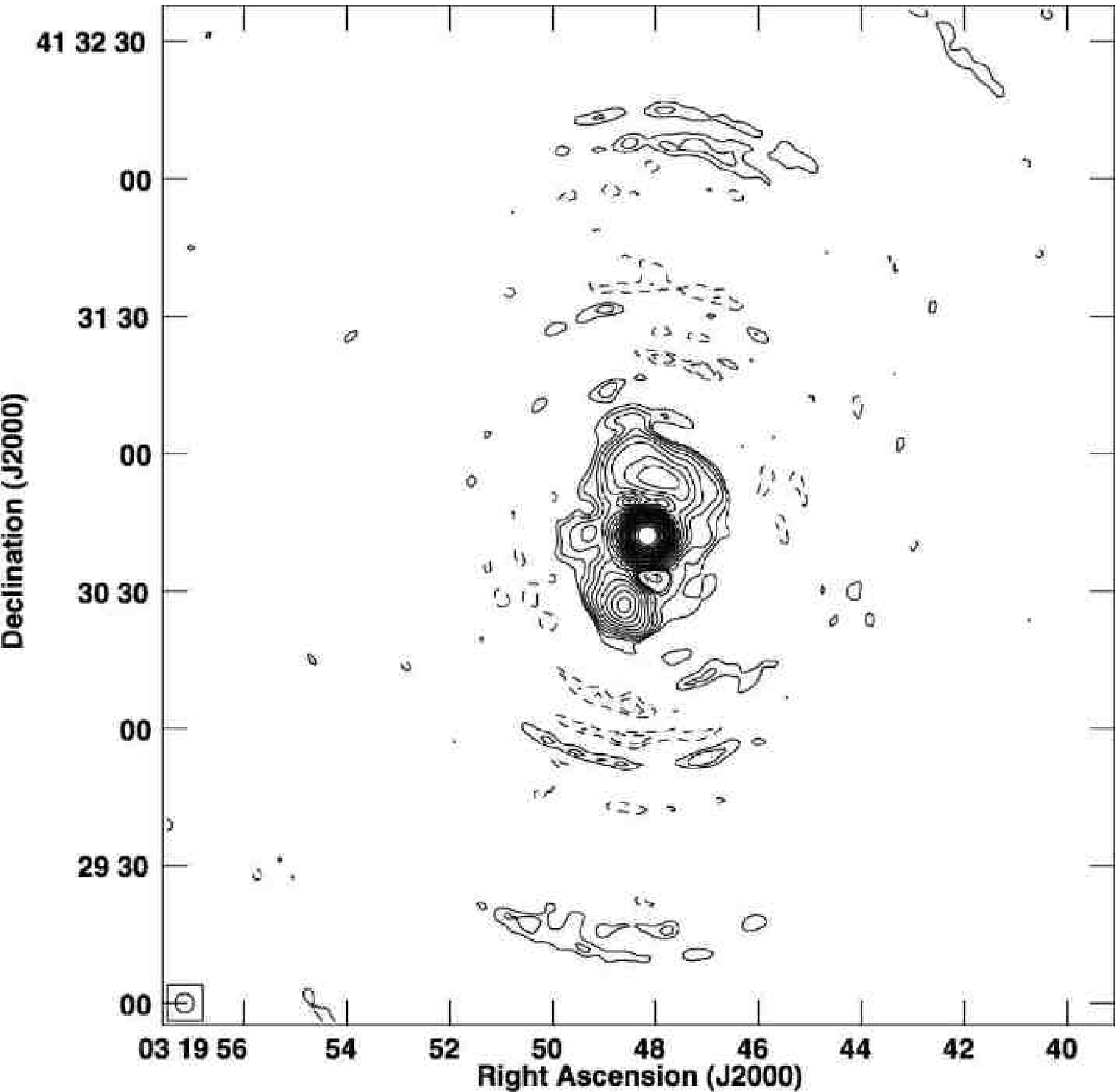}
\centerline{
\includegraphics[height=3.5in]{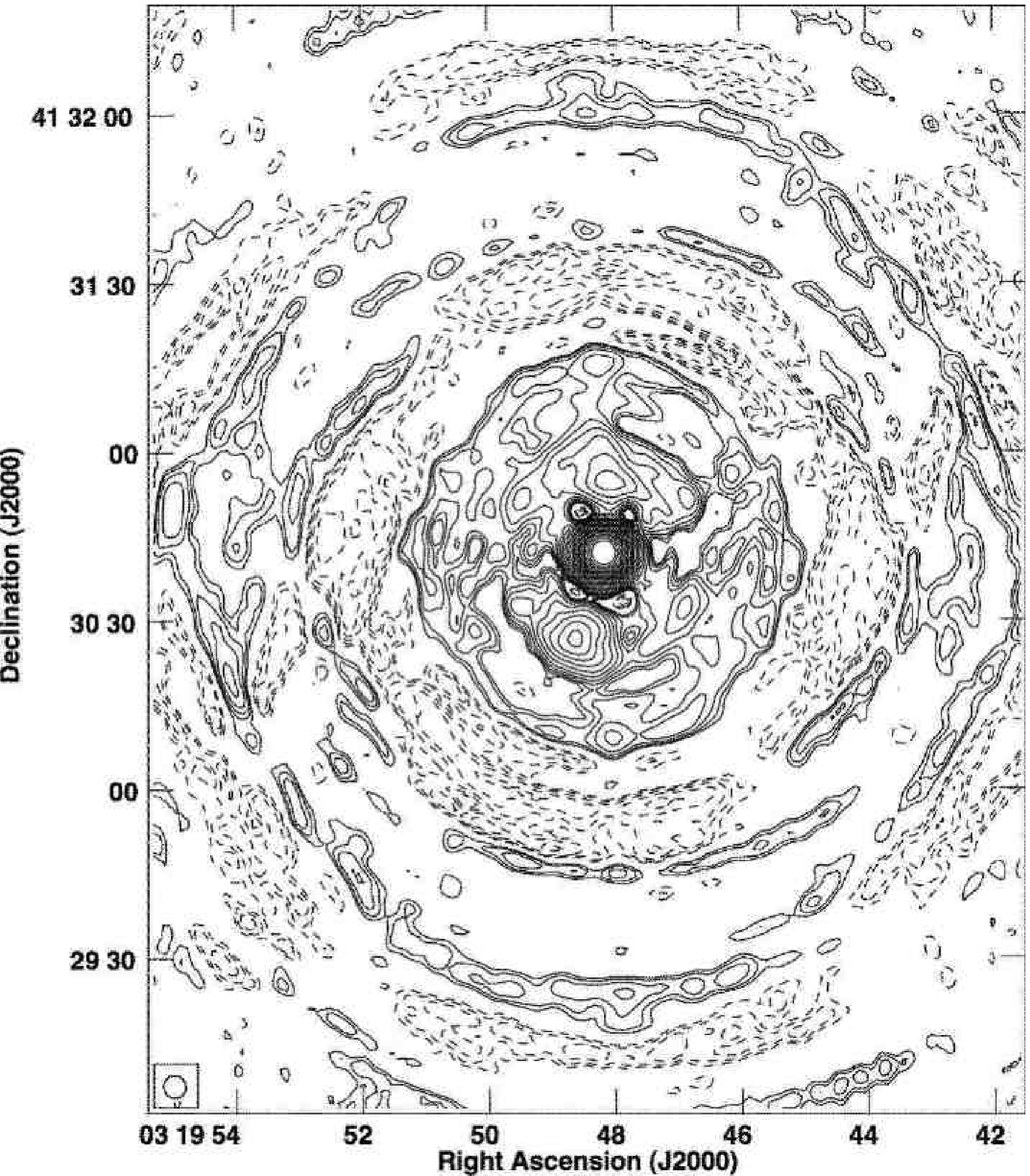}
\includegraphics[height=3.5in]{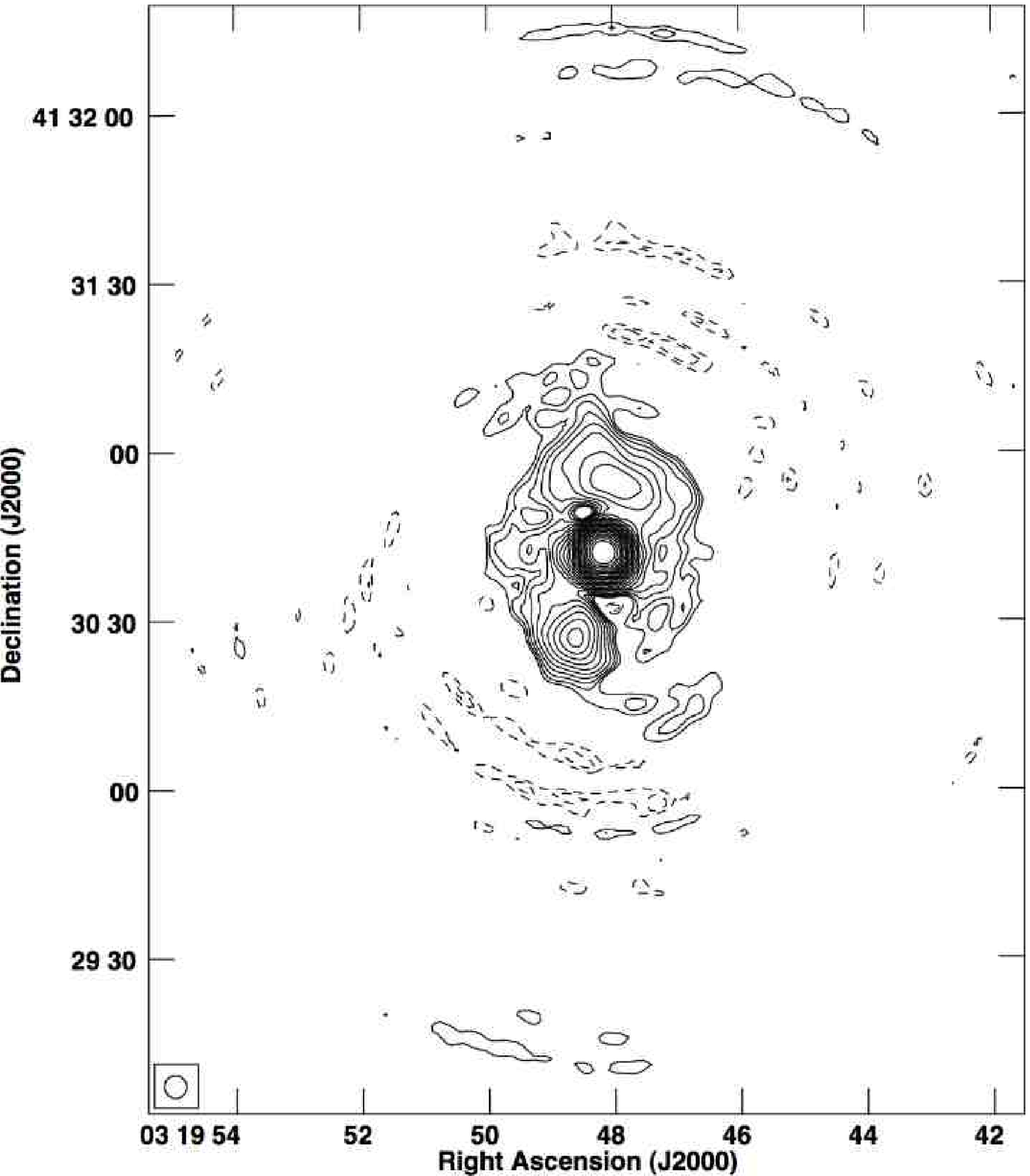}
}
\caption{ 
3C84 with contour interval of powers of $\sqrt{2}$.
The same contouring relative to the peak in the image is used for all
plots.
\hfil\break
{\bf Top:} 3C84 observed at the pointing center.
The most negative contour (dashed) is at -8 mJy/beam
\hfil\break
{\bf Bottom Left:} Portion of wide-field image with 3C84 at the half
power of the antenna pattern.
The most negative contour (dashed) is at -11 mJy/beam
\hfil\break
{\bf Bottom Right:} Like Left except using the autoCenter technique.
The most negative contour (dashed) is at -4 mJy/beam
}
\label{3C84}
\end{figure*}

A test using real data and wide-field imaging was made using the VLA at
1.4 GHz and observations of 3C84 (peak = 24.6 Jy).
3C84 was observed both at the pointing center and at the half power point
of the beam.
The observations were made in spectral mode with 15 channels of 390
kHz bandwidth.
The data were bandpass calibrated in addition to the amplitude and
phase calibration and the edge channels were excluded from processing.
Wide--field 3671$\times$3671 pixel images were made to cover the primary
beam of the antennas using the Fly's Eye technique and a 37 facet mosaic.
In all cases, Obit task Imager {\reffont determined and}
applied amplitude and phase self--calibration to
optimize the dynamic range.
The data set with 3C84 at the half power of the antenna power pattern
was imaged both with and without the autoCenter technique.
In these images, 3C84 was half--way to the edge of its facet
(co--planarity limit) and 0.4 of a cell from the nearest grid cell in
Right Ascension and 0.3 of a cell in Declination.
Sampling was 4 pixels per beam.
Contour plots of a portion of the images around 3C84 are shown in
Figure \ref{3C84}.

\begin{table*}[t]
\caption{3C84 Dynamic Range}
\begin{center}
\begin{tabular}{|l|r|r|r|r|r|} 
\hline\hline
 Image &  Peak       & Near RMS & Far RMS & Near DR & Far DR \\
       &  Jy         & Jy       & Jy      &         &  \\
\hline
center$^1$    & 24.6 & 0.0020 & 0.00122 & 12289 & 20145 \\
half$^2$      & 12.0 & 0.0024 & 0.00121 &  5029 &  9933 \\
half/auto$^3$ & 11.6 & 0.0010 & 0.00055 & 11745 & 21163 \\
\hline
\end{tabular}
\end{center}
\hfill\break
Notes:\hfill\break
$^1$ 3C84 observed at pointing center,\hfill\break
$^2$ 3C84 observed at half power,\hfill\break
$^3$  3C84 observed at half power using autoCenter.
\label{3C84DR}
\end{table*}

It is immediately obvious from Figure \ref{3C84} that the autoCenter
technique helps improve the dynamic range of the image with 3C84 well
away from the pointing center.
This is explored further in Table \ref{3C84DR} which gives the
relevant image statistics.
The RMS was determined in 601$\times$601 pixel windows either
centered on 3C84 (``Near RMS'') or far from 3C84 (``Far RMS'').
The near RMS values were determined using a histogram analysis.

As can be seen from Table \ref{3C84DR}, using the autoCenter technique
doubles the effective dynamic range of the image, even in regions far
from the obvious artifacts caused by 3C84 and does even better near
the source.
The autoCenter image has comparable dynamic range to the observation
with 3C84 on axis.
Especially in the neighborhood of 3C84, it is clear that other
systematics {\reffont are limiting the dynamic range. Indeed, given the extended
emission surrounding 3C84 the images are limited by uv-coverage as the
test observations lacked the necessary short spacings.  We find that this
limitation is more severe than other systematic effects such as bandpass
mismatches, pointing errors and beam squint.  We have re-imaged
the data after discarding baselines shorter than 12~k$\lambda$ and we
have obtained a dynamic range that is $\sim 20$\% higher (again using the
autoCenter technique) although the extended flux that surrounds the core
of 3C84 is, of course, lost to the baseline restriction.}

{\reffonta The difference in execution times for processing with and
without the autoCenter technique depends on the details, i.e., number of
self--calibrations, structure in the field, etc.; but is seldom
significant. 
The cost of the extra, shallow CLEAN to locate sources to be
re-centered can be partially or totally compensated by less time spent
modeling artifacts in subsequent CLEANs.
In the test presented in this section, processing without
autoCentering actually took 1\% longer than with autoCentering.
}

\section{Conclusion}

We have demonstrated that the dynamic range obtained
from imaging interferometric observations can be adversely affected by
pixelization of the images.
Image pixelization effects can limit dynamic range to about $10^4$
even for point sources and non co--planar effects can limit the dynamic
range to about $10^3$ even with a moderately conservative limit on
facet size. 
The higher dynamic range needed for the EVLA and LOFAR, where wide
fields of view with numerous bright sources will be common, need improved
techniques to circumvent these limits.

The autoCenter technique presented here has shown factors of 50 to 1000
improvement in images made from simulated data with no noise or
systematic errors. 
An improvement greater than a factor of 2 was achieved in 
images made from real observations
of the bright source 3C84, even far from the obvious artifacts due to
the source.
{\reffonta Due to the lower level of artifacts, processing using this
technique on the 3C84 test presented above used marginally less
computer time than without. 
}

Fractional pixel corrections are more difficult to implement in a
single image w-projection method.
A simple shift can center a single source, also varying the pixel spacing
could center two sources and a rotation could add a third but
centering a larger number of sources will not be possible.
The introduction of separate, simultaneous ``w--projection'' grids to
accommodate such strong sources would seem to reduce the benefits of
the ``w--projection'' technique\footnote{Alternatively,
a finite number of tangent facets could be
used to supplement the ``w--projection'' grid and ``unboxes'' used in
a manner similar to the implementation discussed here.  Again, this
would seem to reduce the benefits of the ``w--projection'' technique.}.
{\reffont Even grid-less subtraction of strong sources will require some shifting of
their cataloged position due to foreground and (variable) instrumental
effects.  One such procedure discussed in the literature \citep{Voronkov04}
seems to be much more computationally expensive than the procedure
discussed in this paper, even when dealing with a small number of sources.}

The technique described here can be applied to an arbitrary number of
point--like sources.
The tests presented here suggest that high dynamic range imaging of
bright extended sources needs a better set of basis functions than the
delta functions on grid cells that are used by CLEAN.

\begin{acknowledgements}
The authors would like to thank Huib Intema of Leiden University for
stimulating discussions on this subject.
We would also like to thank the anonymous referee for useful comments
leading to an improved presentation of this technique.
\end{acknowledgements}

\bibliographystyle{aa} 
\bibliography{9104}
\end{document}